\documentclass[a4paper]{jpconf}
\usepackage{graphicx}
\begin{document}
\title{Resonance $d^*(2380)$ and higher isospin states}

\author{E.A.Doroshkevich}

\address{Institute for Nuclear Research, RAS,117312, Moscow, Russia}
{\rm ~~~~~~~~~~~~~~~~~~~~For the WASA-at-COSY collaboration}
\ead{eugene.doroshkevich@gmail.com}

\begin{abstract}
 A new isoscalar state in the two-baryon system with mass 2380 MeV and width 80 MeV  denoted now $d^*(2380)$
 has been detected in the experiments at the Juelich Cooler Synchrotron (COSY). Existence or influence of states 
 with higher isospin is a subject of the experiment for studing of the $\pi^+\pi^-$ production in $pp$ collisions.
 Calculations taking into account isotensor dibaryon resonance in the $\Delta N$ with $I(J^p) = 2(1^+)$ provide 
 a good description of the $pp\to pp\pi^+\pi^-$ total and differential cross sections.
\end{abstract}

\section{Observation of the dibaryonic resonances}
The dibaryon resonance $d^*(2380)$ with  $I(J^p ) = 0(3^+)$ has been observed in the reaction 
$pn\to d\pi^0\pi^0$ \cite{16, 20, 21} in the quasi-free $pd$ scattering. 
Since it was shown that resonance features containing contributions of isoscalar components are observed 
in the reactions  $pn\to d\pi^+\pi^-$ \cite{22},  $pn\to pp\pi^0\pi^-$ \cite{23}, $np\to np\pi^0\pi^0$ \cite{24} and  
$pn\to pn\pi^+\pi^-$ \cite{25, 27} effects of high isospin states were measured in the reaction of 
$\pi^+\pi^-$ production.
 Exclusive and kinematically complete measurements of the total and differential cross sections in the reaction 
$pp\to pp\pi^+\pi^-$ were carried out.
It is interesting that $d^*$ corresponds to one of the predicted 1964 by Dyson and Xuong \cite{35} multiplet states and later it discussed in some theoretical investigations \cite{51, 52, 53}.
 
 The theoretical model based on the t-channel meson exchange process including the excitation and decay of 
 the intermediate states of $N^*(1440)$ and $\Delta(1232)\Delta(1232)$ \cite{31, 32} approximately describes 
 the experimental data in the energy region around 1 GeV. For energies above 1 GeV the strength of the
 $N^*(1440)$ excitation and $\rho$ exchange contribution had to be strongly reduced \cite{15} (modified
 "Valencia model"). The renewed model which takes into account the resonance $d^*(2380)$ describes isoscalar
 data. 

 The measurements of the reaction  $pp\to pp\pi^+\pi^-$  were carried out for the quasi-free $pd$ scattering at
 accelerator COSY  (Forschungszentrum Jülich) with WASA detector \cite{39, 40}. Details of the experimental
 setup and data analysis are described in \cite{38, 38a}. The dependence of cross sections on initial energy is
 shown in Fig.1. The dashed curve represents estimations of the modified "Valencia model". 
 One can see that model calculations agree very good at low energies and underestimate experimental results
 at $T_p$>0.9 GeV. Calculations taking into account the isospin relations represented by the shaded band also
 underestimate the cross section mesurements  at $T_p$>0.9 GeV.
 
 Differential distributions contain additional information concerning the process under investigation.
 The invariant mass distributions  $M_{pp}$,  $M_{p\pi^+}$, $M_{pp\pi^+}$,  $M_{\pi^+\pi^-}$,  $M_{p\pi^-}$,
  $M_{pp\pi^-}$ and single particle distributions of polar angle in the center-of-mass $\Theta^{c.m.}_{p}$,
  $\Theta^{c.m.}_{\pi^+}$, $\Theta^{c.m.}_{\pi^-}$ are shown in Fig.2.
 Presented data are efficiency and acceptance corrected. In all plots shaded areas denote the phase space
 distributions.  All of the invariant mass distributions differ from the pure phase space. 

\begin{figure}[t] 
\includegraphics[clip=true,width=0.65\columnwidth,angle=0.]{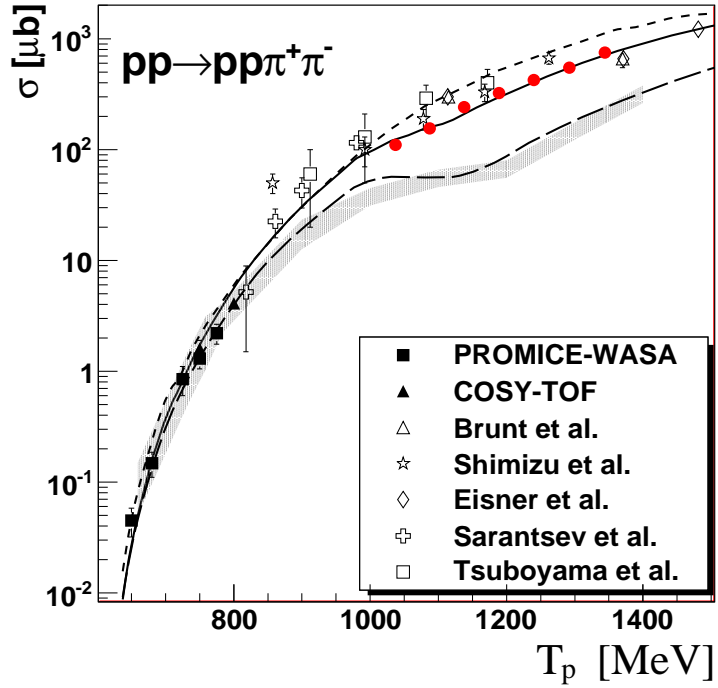}
\caption{\label{T1} \footnotesize   Dependence of the total cross sections on incident proton energy $T_p$ for
 the reaction $pp\to pp\pi^+\pi^-$. The solid circles show results from \cite{38}. Other symbols denote results from
previous measurements \cite{3, 3a, 4, 5, 5a, 5b, 5c, 5d, 5f, 14, 43}. 
 The shaded band displays the isospin-based prediction. The dotted line gives the 
 original Valencia calculation \cite{31}, the dashed line gives the 
"modiﬁed Valencia" calculation [9]. The solid line is obtained, if an associatedly produced $D_{21}$ 
resonance is added according to the process $ pp \to D_{21}\pi^- \to pp\pi^+\pi^-$ with a 
strength adjusted to the total cross section data.}
\end{figure}

\newpage
\begin{figure}[h]
\includegraphics[clip=true,width=0.29\columnwidth,angle=0.]{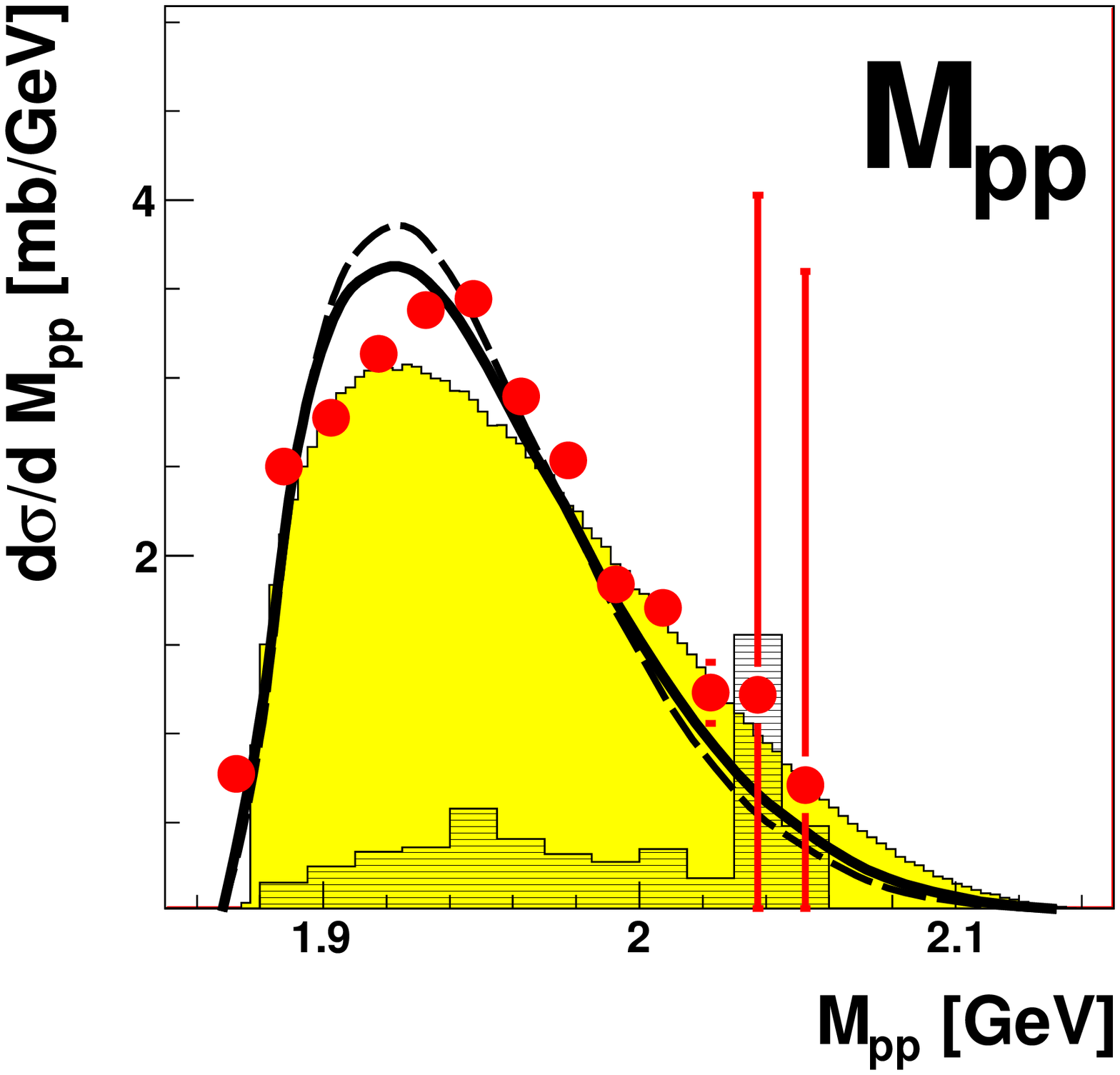}
\includegraphics[clip=true,width=0.29\columnwidth,angle=0.]{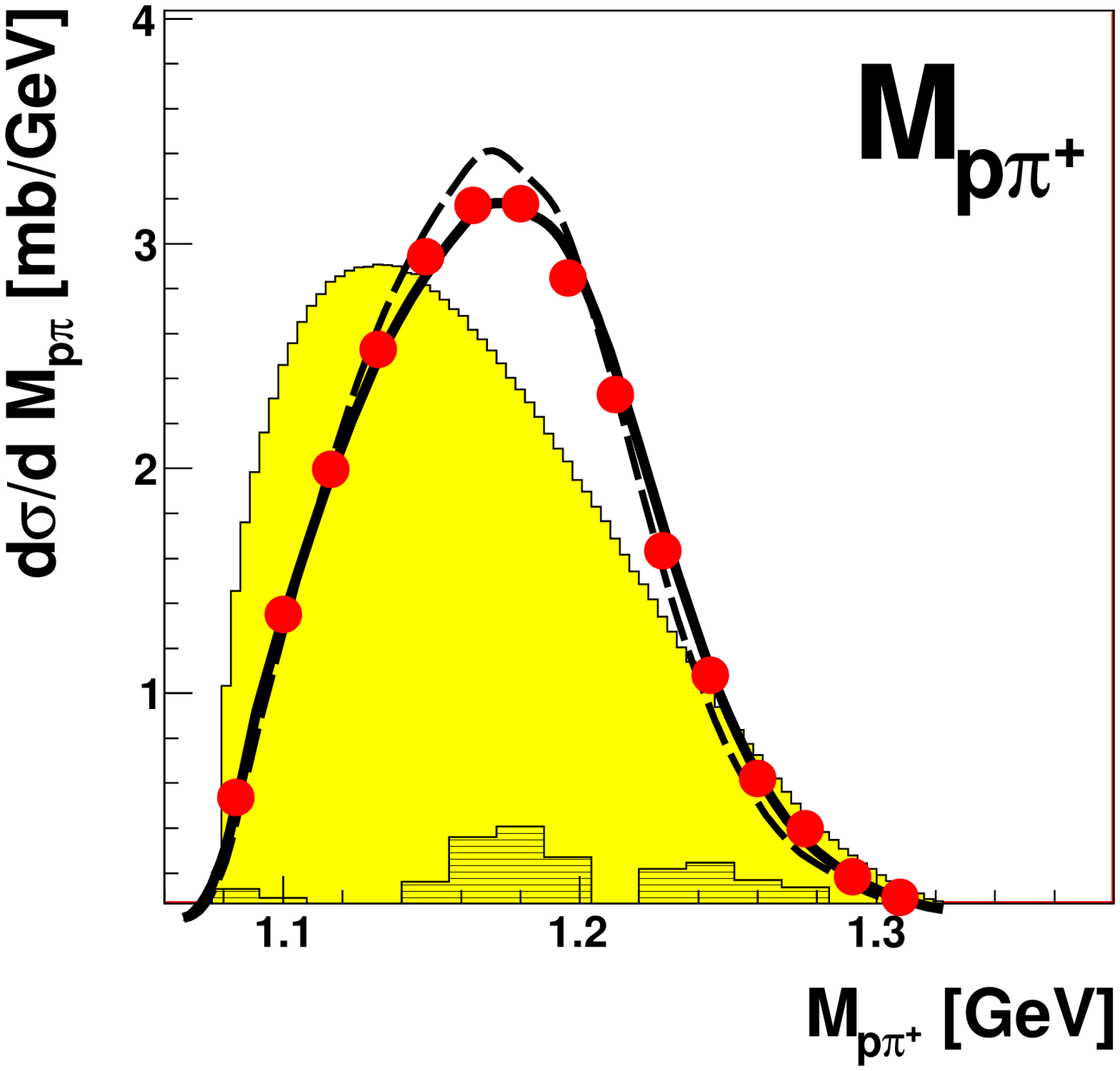}
\includegraphics[clip=true,width=0.29\columnwidth,angle=0.]{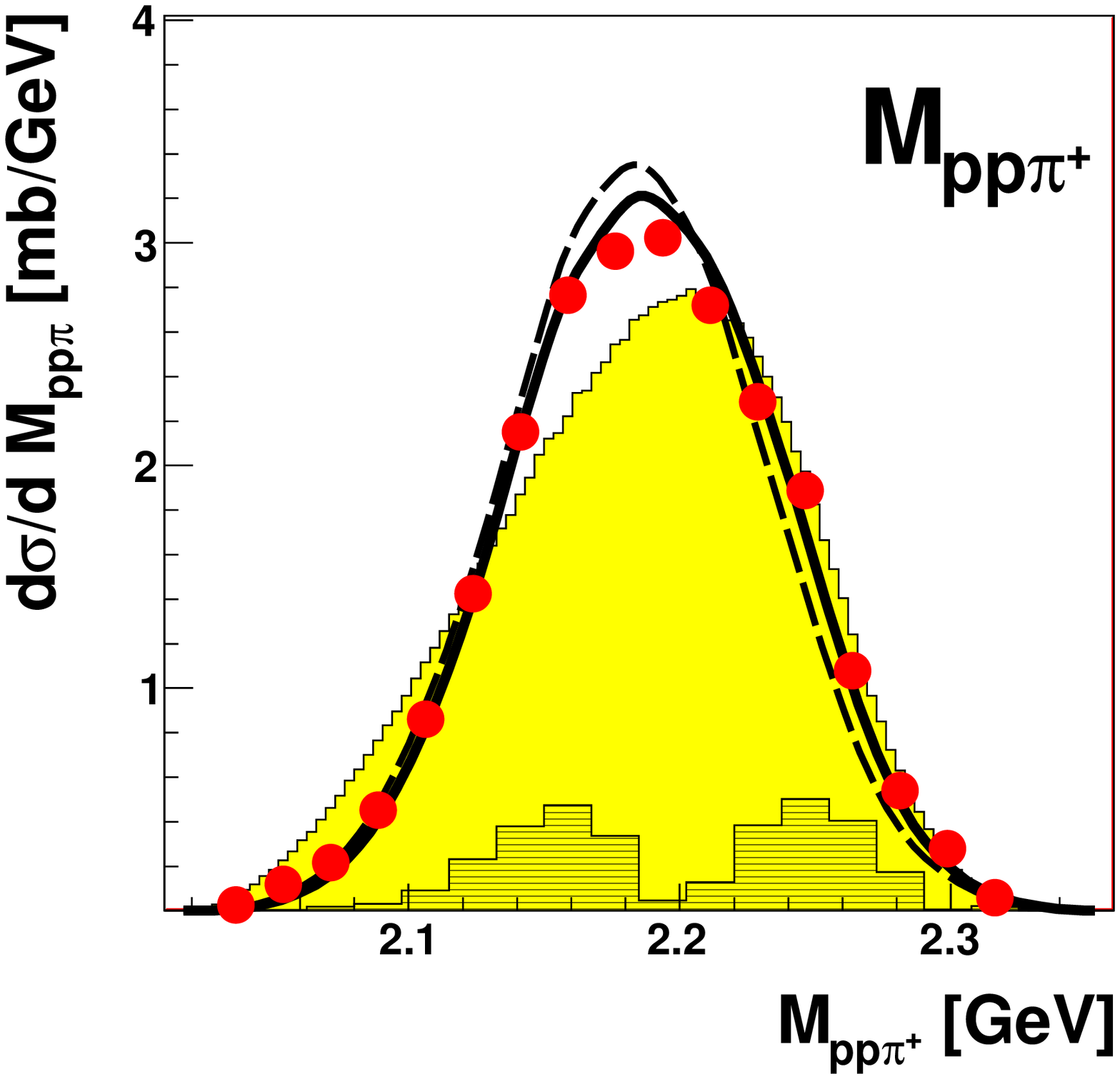}
\end{figure}
\begin{figure}[h]
\includegraphics[clip=true,width=0.29\columnwidth,angle=0.]{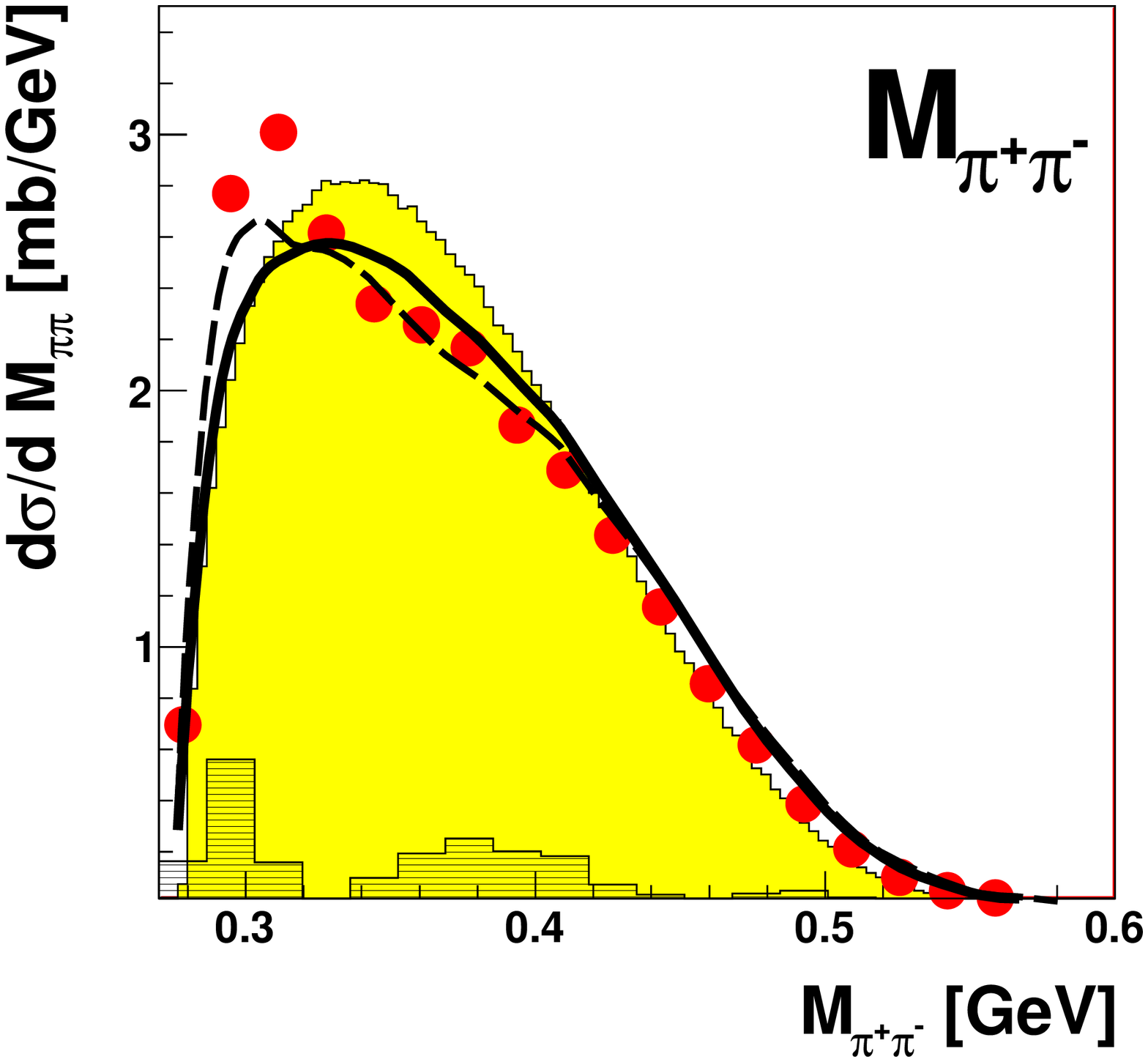}
\includegraphics[clip=true,width=0.29\columnwidth,angle=0.]{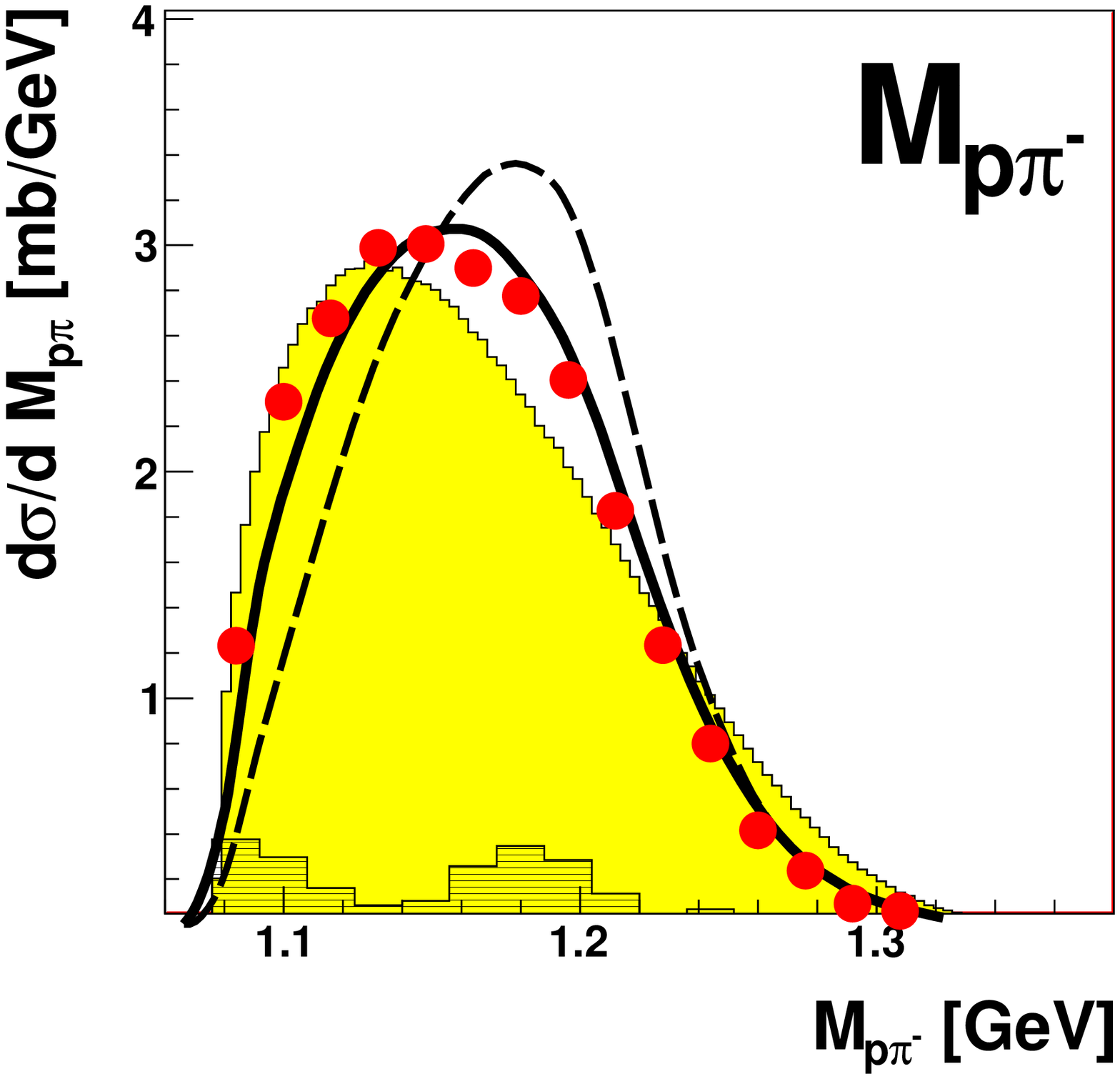}
\includegraphics[clip=true,width=0.29\columnwidth,angle=0.]{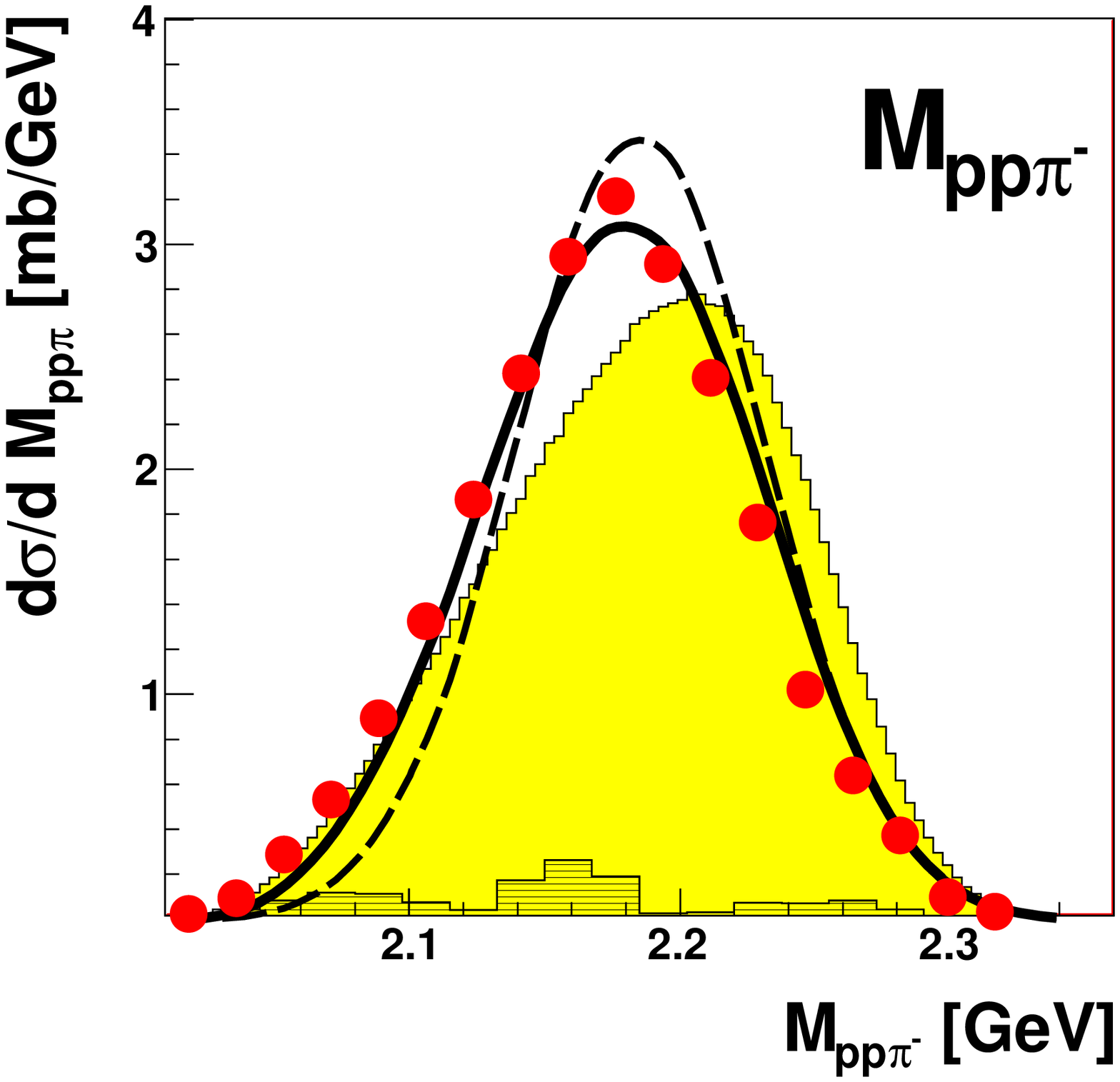}
\end{figure}
\begin{figure}[h]
\includegraphics[clip=true,width=0.29\columnwidth,angle=0.]{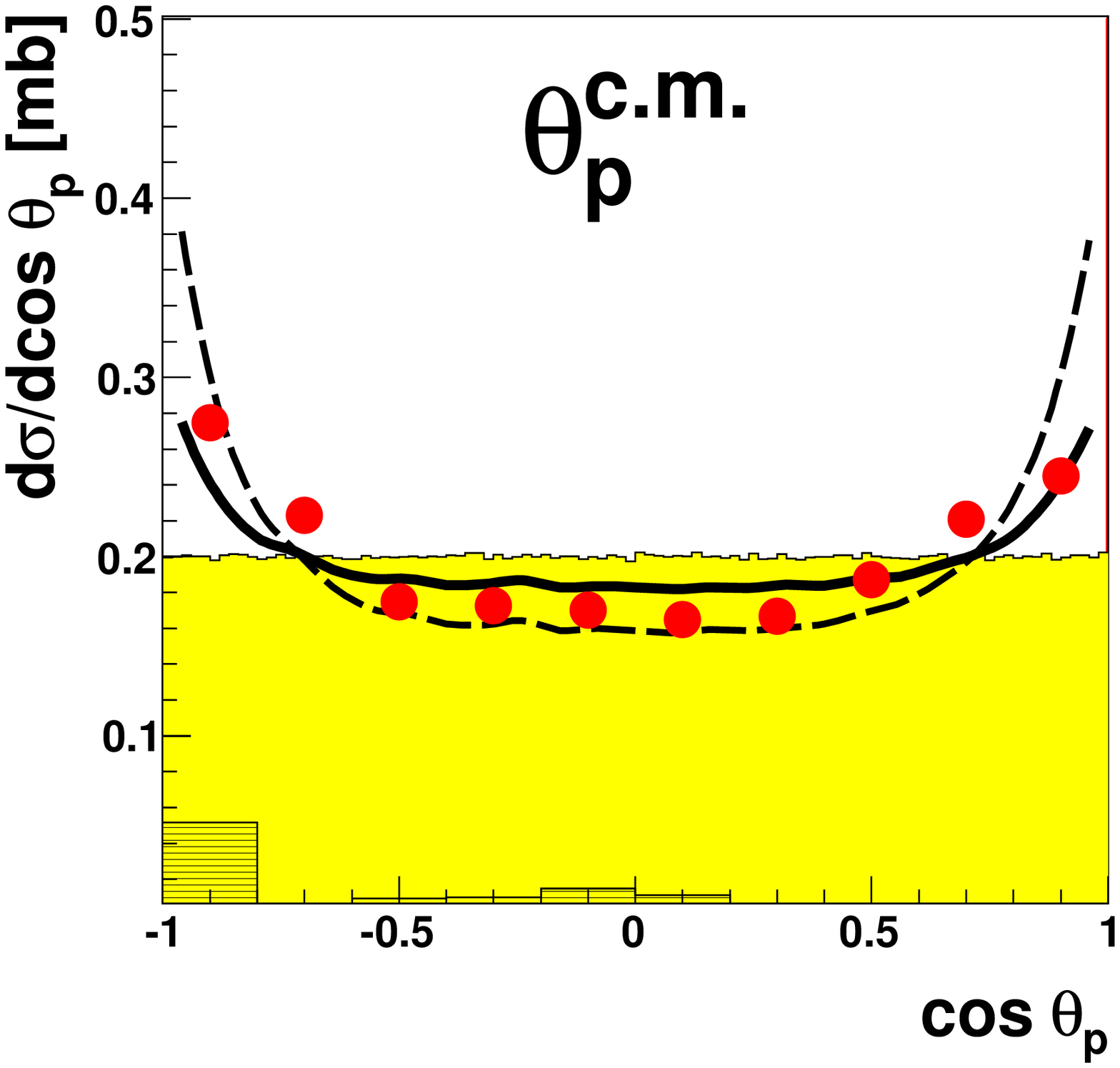}
\includegraphics[clip=true,width=0.29\columnwidth,angle=0.]{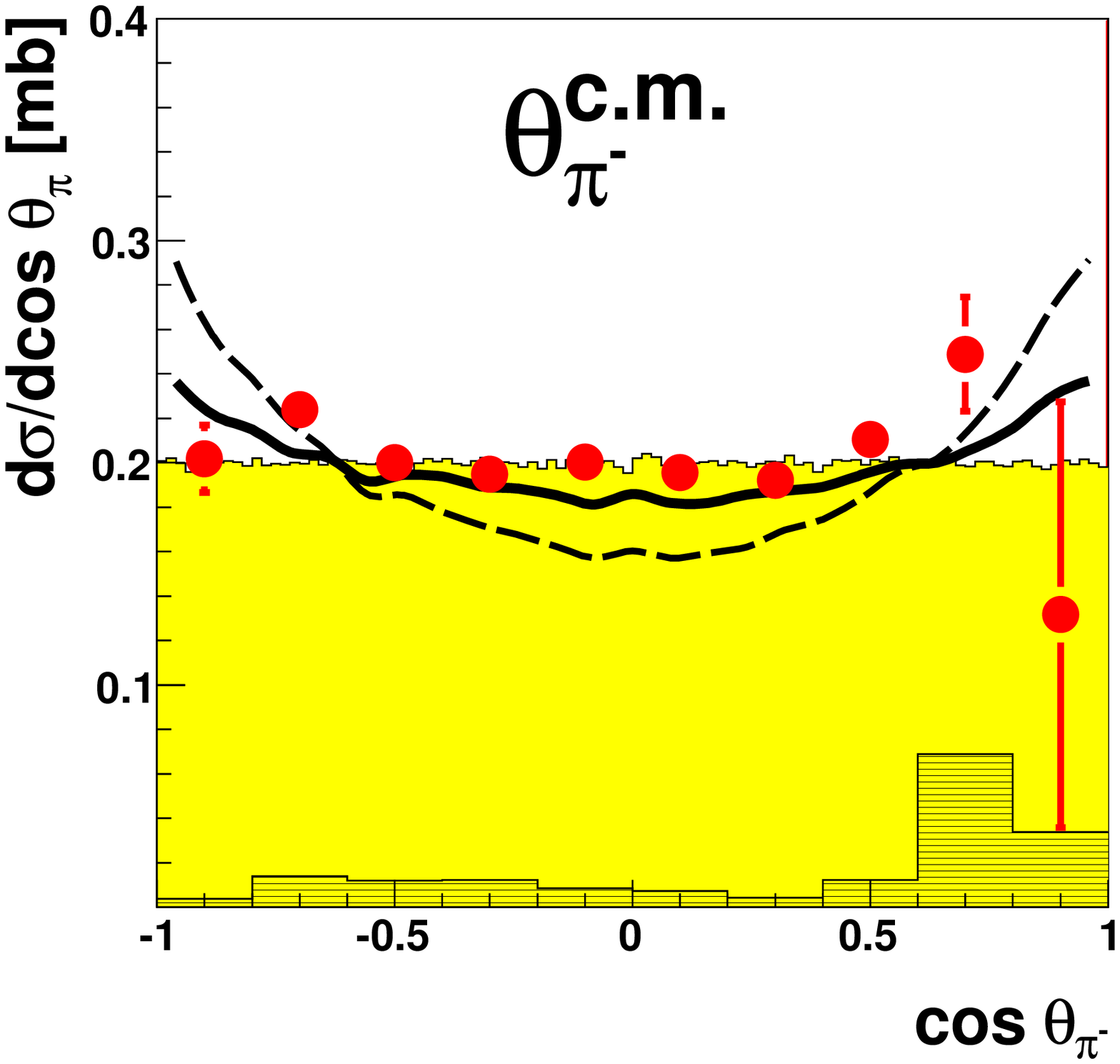}
\includegraphics[clip=true,width=0.29\columnwidth,angle=0.]{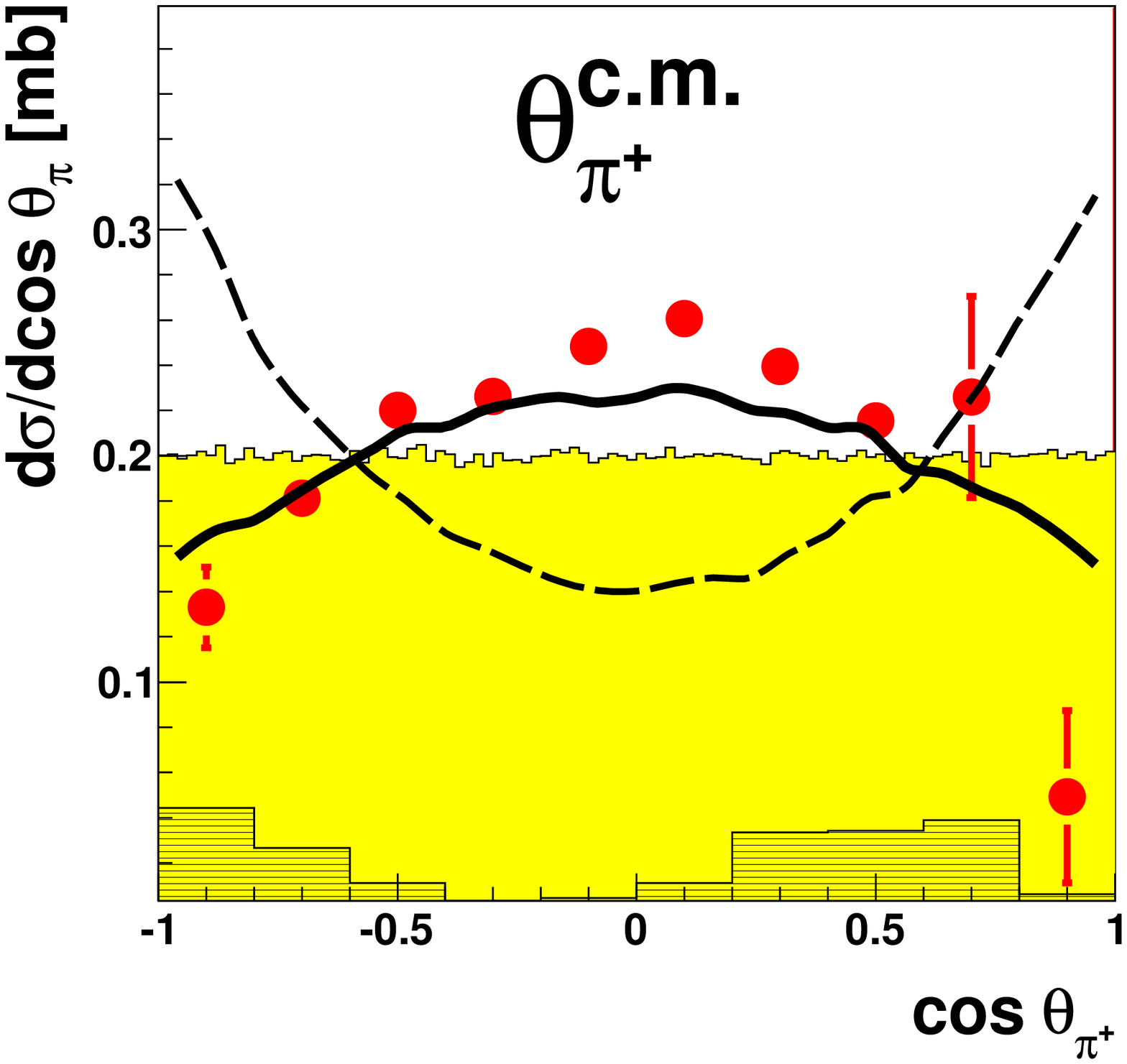}
\caption{\label{WASA} \footnotesize Differential distributions in the region of initial proton energy 
$T_p=1.08-1.36 GeV$ are shown- 
Upper raw: $M_{pp}$, $M{p\pi^+}$ and $M_{pp\pi^+}$; 
Middle raw: $M_{pi^+\pi^-}$, $M_{p\pi^-}$ and $M_{pp\pi^-}$;
 Lower raw: Angular dependence of $p$, $\pi^-$, $\pi^+$ in c.m. \cite{16}.
 Shaded areas (yellow) denote the Phase Space distributions. 
The hatched histograms indicate systematic uncertainties due to the restricted Phase Space coverage of the data.
Dashed curve shows Modified "Valencia model", solid curves are fit with isotensor $D_{21}$ state}
\end{figure}

Calculations of models which are shown by curves are normalized to experimental data with area. 
The modified "Valencia model" calculations are shown by dashed curves. There is difference between
 experimental data and modified "Valencia model" calculations for the distributions $M_{p\pi^-}$ and
  $M_{pp\pi^-}$. For the angular distributions there is striking contrast between $\pi^+$ and $\pi^-$.
 If the $\Delta\Delta$ mechanism is supposed as dominant in t-channel process then the distributions in each pair
 $M_{p\pi^-}$, $M_{p\pi^+}$ and $M_{pp\pi^-}$, $M_{pp\pi^+}$ should be similar. But this is not the case.

Since the modified "Valencia model" mechanism failed in the experimental data description one can suppose
 that predicted by Dyson and Xuong \cite{35} the isotensor $\Delta N$ state $D_{21}$ with mass 2144-2148 MeV and $I(J^p)=2(1^+)$ contributes to this process. Assuming such resonance in the process  $pp\to D_{21}\pi^-\to pp\pi^+\pi^-$ one can compare calculations (solid curve in Figs.2) with data. 

\section{Conclusions}

 Differential distributions $M_{p\pi^+}$,  $M_{pp\pi^+}$ differ from corresponding $M_{p\pi^-}$ and $M_{pp\pi^-}$. Also there is contrast between the angular distributions of $\Theta^{c.m.}_{\pi^+}$ and  $\Theta^{c.m.}_{\pi^-}$. The use of the theoretically predicted isotesor $\Delta N$ state $D_{21}$ provides qualitative agreement in the experimental data description. 

\subsection{Acknowledgments}
Authors wishing to acknowledge the ICPPA-2018 organizing committee.
We are grateful to the WASA-at-COSY collaboration and especially grateful to H.Clement, M.Bashkanov, T.Skorodko.
 We acknowledge valuable discussions with M.Schepkin and Y.Dong.

\end{document}